\documentclass[aps,pre,preprint,groupedaddress,showpacs]{revtex4-1}

\usepackage{graphicx}% Include figure files
\usepackage{amsmath}
\usepackage{amsfonts}
\usepackage{dcolumn}% Align table columns on decimal point
\usepackage{bm}% bold math
\usepackage{mathrsfs}
\usepackage{mathtools}
\begin{document}

%\preprint{}

\title{The turbulence velocity gradient tensor formed additively 
by normal and non-normal tensors}


%\author{}
%\thanks{}
%\affiliation{}
\author{C. J. Keylock}
\affiliation{Sheffield Fluid Mechanics Group and Department of Civil and Structural Engineering, University of Sheffield, Mappin Street, Sheffield, U.K., S1 3JD }%
 \email{c.keylock@sheffield.ac.uk}


\date{\today}

%\begin{abstract}
%We decompose the velocity gradient tensor for turbulence into normal and non-normal parts, and condition our analysis on the strain eigenvector alignments between these tensors. We identify states that always enhance, and always counteract the axisymmetric expansion state, and  give a rationale for decomposing the production balance term into its constituents: complex behavior arises when the dominant strain alignments involve the non-normal tensor. Finally, we develop a topological analysis framework where mathematical bounds on two of the three variables leads to an analysis in two planes.
%\end{abstract}

%\pacs{47.27.Ak,47.27.Gs, 47.10.-g, 02.40.-k}
%\keywords{}

\maketitle

\section{Introduction}
%\subsection{}
%\subsubsection{}
An enhanced understanding of how a turbulent flow dissipates energy is crucial for linking the topological view of turbulence to the statistical considerations of dissipation and, hence, developing the next generation of numerical closure schemes for modeling high Reynolds number flows in industry and the environment. This is an old problem with pioneering work undertaken by G. I. Taylor in the 1930s \cite{taylor37}, and Betchov in the 1950s \cite{betchov56}, and significant progress having been made since the advent of high Reynolds number computations \cite{kerr85, ashurst87,meneveau11}. 

The Navier-Stokes equations can be written in terms of the velocity-gradient tensor
\begin{equation}
A_{ij} =\biggl(\begin{smallmatrix}
\partial u_{1}/\partial x_{1} & \partial u_{1}/\partial x_{2} & \partial u_{1}/ \partial x_{3}\\ 
\partial u_{2}/\partial x_{1} & \partial u_{2}/\partial x_{2} & \partial u_{2}/ \partial x_{3}\\ 
\partial u_{3}/\partial x_{1} & \partial u_{3}/\partial x_{2} & \partial u_{3}/ \partial x_{3}
\end{smallmatrix} \biggr),
\end{equation}
and a topological classification of the flow can be developed using the characteristic equation for $\mathbf{A}$\cite{chong90}: $\lambda^{3} + \mbox{P}\lambda^{2} + \mbox{Q}\lambda +\mbox{R} = 0$ , where $\mbox{P} = -\sum\lambda_{i} = 0$ in an incompressible flow because of the divergence-free constraint, and the $\lambda_{i}$ are the eigenvalues of $\mathbf{A}$. The term $\mbox{Q}$ may be interpreted as the excess of total enstrophy to total strain and has become popular as a means to visualize coherent structures \cite{hunt88,dubief00}. That is, if $\mathbf{A}$ is decomposed into symmetric strain rate and skew-symmetric rotation rate components as $A_{ij} = S_{ij} + \Omega_{ij}$, $\mbox{Q} = -\frac{1}{2} \mbox{tr}(\mathbf{A}^{2}) = \frac{1}{2}\big(\Omega^{2} - S^{2}\big)$, where $\Omega^{2}$ is the total enstrophy and $S^{2}$ is the total strain. Transforming rotation into the vorticity, $\boldsymbol{\omega} = \epsilon_{ijk}\boldsymbol{\Omega}$, where $\epsilon_{ijk}$ is the Levi-Civita symbol, leads to transport equations for these terms:
\begin{eqnarray}
\frac{1}{2} \frac{D S^{2}}{Dt} &=& -S_{ij}S_{jk}S_{ki} - \frac{1}{4}\omega_{i}\omega_{j}S_{ij} \nonumber \\
&-& S_{ij}\frac{\partial^{2} p}{\partial x_{i} \partial x_{j}} + \nu S_{ij}\nabla^{2} S_{ij} \nonumber \\
\frac{1}{2} \frac{D \omega^{2}}{Dt} &=& \omega_{i}\omega_{j}S_{ij} + \nu \omega_{i}\nabla^{2}\omega_{i},
\label{eq.S2Om2}
\end{eqnarray}
The deviatoric part of the pressure Hessian, $-S_{ij}\frac{\partial^{2} p}{\partial x_{i} \partial x_{j}}$, contains important information on the non-local properties of turbulence \cite{ohkitani95,chevillard08,wilczek14}. However, in the mean, its contribution is zero \cite{tsinober01}, meaning that a popular approach to analyzing the evolution of total strain and enstrophy is the reduced Euler framework \cite{cantwell92}. This means that the key terms are the strain rate production, $-S_{ij}S_{jk}S_{ki}$ and enstrophy production, $\omega_{i}\omega_{j}S_{ij}$. These appear in the equation for $\mbox{R}$:
\begin{equation} 
\mbox{R} = - \det \lambda_{i} = \frac{1}{3}\big(-S_{ij}S_{jk}S_{ki} - \frac{3}{4}  \omega_{i}\omega_{j}S_{ij} \big),
\end{equation}
and the discriminant function in $\mbox{Q}-\mbox{R}$ space that separates regions with a conjugate pair of eigenvalues (that act as saddles or nodes) to those where all $\lambda_{i} \in \Re$ (that act as foci) is given by $\mbox{D} = \mbox{Q}^{3} + (27/4)\mbox{R}^{2}$ \cite{perry87}. While studies of the dynamics of the velocity gradient tensor typically analyze the flow in a $\mbox{Q}-\mbox{R}$ space \cite{chong90,tsinober97} it has been argued that this space needs to be expanded into three dimensions because the Lagrangian dynamics in a $\mbox{Q}-S_{ij}S_{jk}S_{ki}-\omega_{i}\omega_{j}S_{ij}$ space reveals novel and complementary properties on the $\mbox{Q}-\omega_{i}\omega_{j}S_{ij}$ plane \cite{luthi09}.

The intention of this paper is to propose a contrasting mathematical starting point for an additive decomposition of $\mathbf{A}$ based on the notion of matrix/tensor normality, and to demonstrate the fluid mechanical relevance of such an approach. This includes some new insights into results for the paradigmatic case of homogeneous, isotropic turbulence.

\section{Tensor normality}
If $\mathbf{A}$ is normal, then 
\begin{equation}
\mathbf{A}\mathbf{A}^{\mbox{H}} = \mathbf{A}^{\mbox{H}}\mathbf{A},
\label{eq.norm}
\end{equation} 
where $\mbox{H}$ is the conjugate transpose. If $\mathbf{A}\mathbf{A}^{\mbox{H}} \ne \mathbf{A}^{\mbox{H}}\mathbf{A}$ then there are some `residual' dynamics that exist independent of the eigenvalue representation of $\mathbf{A}$. This may be made explicit by considering the Schur decomposition of $\mathbf{A}$:
\begin{eqnarray}
\mathbf{U}\mathbf{T}\mathbf{U}^{\mbox{H}} &=& \mathbf{A} \nonumber \\
\mathbf{T} &=& \boldsymbol{\Lambda} + \mathbf{N}\nonumber \\
\mathbf{U}\boldsymbol{\Lambda}\mathbf{U}^{\mbox{H}} &=& \mathbf{B}\nonumber \\
\mathbf{U}\mathbf{N}\mathbf{U}^{\mbox{H}} &=& \mathbf{C},
\label{eq.schur}
\end{eqnarray}
where $\mathbf{U}$ is unitary and the Schur matrix, $\mathbf{T}$, may be decomposed into a diagonal matrix of eigenvalues, $\mbox{diag}(\boldsymbol{\Lambda}) = \lambda_{1}, \ldots, \lambda_{3}$ and an upper triangular matrix, $\mathbf{N}$ that characterizes the non-normality of $\mathbf{A}$ \cite{GolubVanLoan13}. We use a complex Schur transform to ensure $\mathbf{T}$ is triangular, rather than the quasi-triangular real form. While this has the disadvantage that $\mathbf{B}$ and $\mathbf{C}$ are potentially complex, computational uncertainties in moving from complex to real forms are obviated. In practice, we work with the strain rate and rotation rate tensors derived from $\mathbf{B}$ and $\mathbf{C}$, i.e. $\mathbf{S}_{\mathbf{B}} = \frac{1}{2}\big(\mathbf{B} + \mathbf{B}^{\mbox{H}}\big)$; $\boldsymbol{\Omega}_{\mathbf{B}} = \frac{1}{2}\big(\mathbf{B} - \mathbf{B}^{\mbox{H}}\big)$, which side-steps this issue. As with the traditional approach, $\mathbf{A} = \mathbf{S}_{\mathbf{A}} + \boldsymbol{\Omega}_{\mathbf{A}}$, our decomposition is additive ($\mathbf{A} = \mathbf{B} + \mathbf{C}$).

An alternative definition of non-normality to (\ref{eq.norm}) is $||\mathbf{A}||_{F}^{2} - ||\boldsymbol{\Lambda}||_{F}^{2} = ||\mathbf{N}||_{F}^{2}$, meaning that the Frobenius norm, $||\ldots||_{F} = \sqrt{\mbox{tr}(\mathbf{A}\mathbf{A}^{\mbox{H}})}$ is a logical choice for this problem. It follows immediately that $||\mathbf{A}||_{F}^{2} = ||\mathbf{B}||_{F}^{2} + ||\mathbf{C}||_{F}^{2}$, $||\mathbf{S}_{\mathbf{A}}||_{F}^{2} = ||\mathbf{S}_{\mathbf{B}}||_{F}^{2} + ||\mathbf{S}_{\mathbf{C}}||_{F}^{2}$, and $||\boldsymbol{\Omega}_{\mathbf{A}}||_{F}^{2} = ||\boldsymbol{\Omega}_{\mathbf{B}}||_{F}^{2}|| + ||\boldsymbol{\Omega}_{\mathbf{C}}||_{F}^{2}$, while $\sum \lambda_{i}^{(\mathbf{C})} = 0$ means that $||\mathbf{S}_{\mathbf{C}}||_{F} = 
||\boldsymbol{\Omega}_{\mathbf{C}}||_{F}$. As a consequence, we may write that
\begin{equation}
||\mathbf{A}||_{F}^{2} = ||\mathbf{S}_{\mathbf{B}}||_{F}^{2} + 2 ||\mathbf{S}_{\mathbf{C}}||_{F}^{2} + ||\boldsymbol{\Omega}_{\mathbf{B}}||_{F}^2.
\label{eq.Arelation}
\end{equation}
This leads us to two indices characterizing the ratio of the norms for the straining parts of $\mathbf{B}$ and $\mathbf{C}$, and the ratio of the straining and rotational norms of $\mathbf{B}$, respectively:
\begin{eqnarray}
\kappa_{1} &=& \frac{||\mathbf{S}_{\mathbf{B}}||_{F}^{2} - 2 ||\mathbf{S}_{\mathbf{C}}||_{F}^{2}}{||\mathbf{S}_{\mathbf{B}}||_{F}^{2} + 2 ||\mathbf{S}_{\mathbf{C}}||_{F}^{2}} \nonumber \\
\kappa_{2} &=& \frac{||\boldsymbol{\Omega}_{\mathbf{B}}||_{F}^{2} - ||\mathbf{S}_{\mathbf{B}}||_{F}^{2}}{||\boldsymbol{\Omega}_{\mathbf{B}}||_{F}^{2} + ||\mathbf{S}_{\mathbf{B}}||_{F}^{2}}.
\label{eq.kappa12}
\end{eqnarray}
To these, we add a third measure that examines the straining properties of $\mathbf{B}$. That is, we apply the Lund and Rogers normalization \cite{LR94} to $\mathbf{B}$ rather than its standard application to $\mathbf{A}$:
\begin{equation}
\kappa_{3} \equiv e_{LR}^{(\mathbf{B})} = \frac{3\sqrt{6}\mbox{R}_{\mathbf{S}}^{(\mathbf{B})}}{\left(-2\mbox{Q}_{\mathbf{S}}^{(\mathbf{B})}\right)^{\frac{3}{2}}}.
\label{eq.kappa3}
\end{equation}

\section{Results}
This study makes use of the Johns Hopkins numerical simulation of HIT at a Taylor Reynolds number of 433 \cite{yili}, which has become a popular resource for studying flow topologies \cite{wan2010,lawsondawson15}. Here we interrogated a $1024^{3}$ volume of the database at one point in time, a strategy that has been adopted previously \cite{wilczek14}. We first show some of the physics that can be uncovered with our approach, before exploring some of the properties of the $\kappa$ planes. Our analysis is conditioned on alignments of the strain eigenvectors, as well as locations in $\mbox{Q}-\mbox{R}$ space.

\begin{figure}[t]
\vspace*{2mm}
\begin{center}
\includegraphics[width=14.2cm]{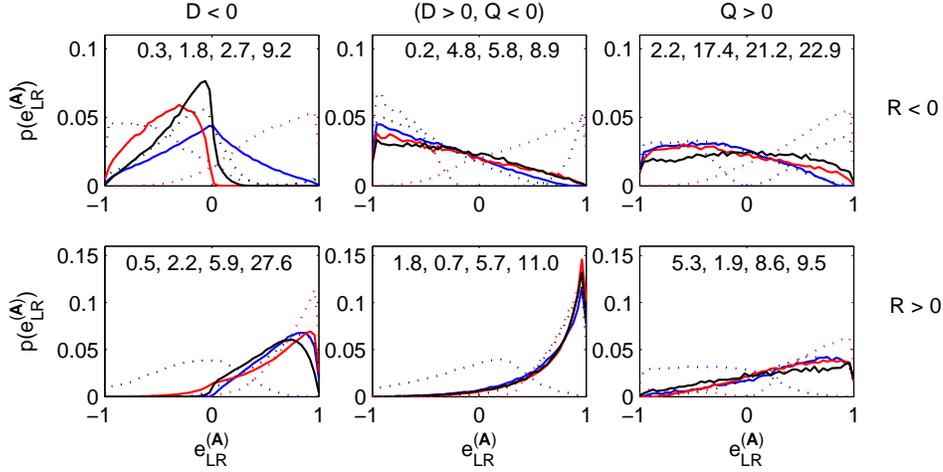}
\end{center}
\caption{Histograms of the Lund and Rogers normalization for the second eigenvalue of the strain tensor of $\mathbf{A}$ as a function of $\mbox{Q}$, $\mbox{D}$, $\mbox{R}$, and the dominant alignment between strain eigenvectors given by $\mbox{max}\big[\cos(\mathbf{e}_{i}^{(\mathbf{A})},\mathbf{e}_{i}^{(\mathbf{B},\mathbf{C})})\big] > 0.9397$, where $i = 1$ are in blue, $i = 2$ in black, and $i = 3$ in red. Alignments between $\mathbf{e}_{i}^{(\mathbf{A})}$ and $\mathbf{e}_{i}^{(\mathbf{B})}$ are displayed as solid lines, and $\mathbf{e}_{i}^{(\mathbf{A})}$ and $\mathbf{e}_{i}^{(\mathbf{C})}$ as dotted lines. Four numbers are quoted in each panel. From left to right these are the percentage occurrence in a given panel for $\big [\mathbf{e}_{1}^{(\mathbf{A})}, \mathbf{e}_{1}^{(\mathbf{C})} \big ]$, (blue, dotted lines), $\big [\mathbf{e}_{3}^{(\mathbf{A})}, \mathbf{e}_{3}^{(\mathbf{C})}\big ]$ (red, dotted lines), $\sum \big [\mathbf{e}_{i}^{(\mathbf{A})}, \mathbf{e}_{i}^{(\mathbf{C})}\big ]$ (all dotted lines), and all the displayed results.}
\label{fig.eLR_A}
\end{figure}

\subsection{Properties of the second eigenvector of the strain tensor}
An important property of homogeneous, isotropic turbulence (HIT) is the tendency for the Lund and Rogers normalization of $\mathbf{A}$ to give values close to +1 as a consequence of preferential axisymmetric extension \cite{meneveau11}. Here we examine this quantity, $e_{LR}^{(\mathbf{A})}$, using the Schur decomposition to partition the results by strain eigenvector alignments as shown in Fig. \ref{fig.eLR_A}. The six dominant strain eigenvector alignments shown in each panel account for 89\% of the total data (i.e. the sum of the right-hand numbers in each panel). This was because in 5\% of cases, there was no observed alignment where $\cos(\mathbf{e}_{i}^{(\mathbf{A})},\mathbf{e}_{i}^{(\mathbf{B},\mathbf{C})}) > 0.9397$ (i.e. $\pm 20$ degrees from perfect alignment), and in 6\% of cases, the strongest alignment was not between strain eigenvectors of the same rank order, i.e. $\mbox{max}\big[ \cos(\mathbf{e}_{i}^{(\mathbf{A})},\mathbf{e}_{j}^{(\mathbf{B},\mathbf{C})})\big] > 0.9397$, but $i \ne j$. 

When $\mbox{Q} > 0$, dominant alignments are between $\mathbf{S}_{\mathbf{A}}$ and $\mathbf{S}_{\mathbf{C}}$ (93\% of cases for $\mbox{R} < 0$ and 91\% for $\mbox{R} > 0$), and the proportion of tensors where this is true for $\mbox{R} < 0$ always exceeds that for $\mbox{R} > 0$ (65\% compared to 52\% when $\mbox{D} > 0$, $\mbox{Q} < 0$, and 29\% compared to 21\% for $\mbox{D} < 0$). Furthermore, a dominant alignment between $\big [\mathbf{e}_{1}^{(\mathbf{A})}, \mathbf{e}_{1}^{(\mathbf{C})}\big ]$ always detracts from the +1 alignment, while $\big [\mathbf{e}_{3}^{(\mathbf{A})}, \mathbf{e}_{3}^{(\mathbf{C})}\big ]$ promotes it. The proportion of these occurrences is given by the numbers quoted on the left and second from the left, respectively, in each panel. In contrast to the alignments between $\mathbf{S}_{\mathbf{A}}$ and $\mathbf{S}_{\mathbf{C}}$, those between $\mathbf{S}_{\mathbf{A}}$ and $\mathbf{S}_{\mathbf{B}}$, generally exhibit a similar behavior to each other. Exceptions include the $\mbox{D} < 0$, $\mbox{R} < 0$ region (10\% of all cases), where $\big [\mathbf{e}_{3}^{(\mathbf{A})}, \mathbf{e}_{3}^{(\mathbf{B})}\big ]$ alignments detract from the $e_{LR}^{(\mathbf{A})} \sim +1$ relation to a greater extent, and where $\big [\mathbf{e}_{1}^{(\mathbf{A})}, \mathbf{e}_{1}^{(\mathbf{B})}\big ]$ (the dominant alignment here, representing 4.6\% of all cases) gives a clear preference for a plane shear configuration of the velocity gradient tensor, $e_{LR}^{(\mathbf{A})} \sim 0$. 

\begin{figure}[t]
\vspace*{2mm}
\begin{center}
\includegraphics[width=14.2cm]{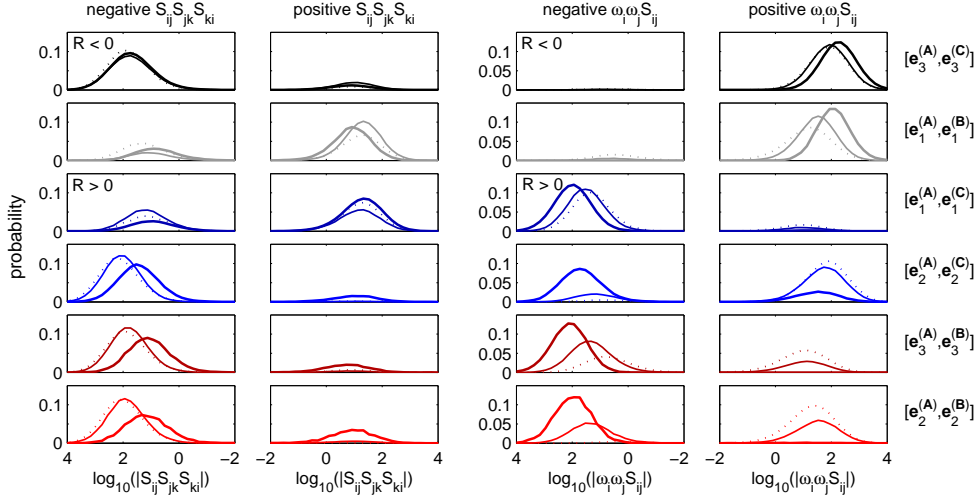}
\end{center}
\caption{Histograms of selected distribution functions, as discussed in the text, for strain production, $S_{ij}S_{jk}S_{ki}$, and enstrophy production $\omega_{i}\omega_{j}S_{ij}$. The abscissa is on a log-scale and positive and negative contributions have been separated. In each panel, the dotted line is for the $\mbox{D} < 0$ case, the solid line is the $(\mbox{D} > 0, \mbox{Q} < 0)$ case, and the thick solid line is the $\mbox{Q} > 0$ case.}
\label{fig.sss_wws}
\end{figure}

It  is clear that when the flow is close to the Vieillefosse tail ($\mbox{R} > 0$ and $\mbox{Q} < 0$), we find $e_{LR}^{(\mathbf{A})} \sim 1$, with the exception of $\big [\mathbf{e}_{1}^{(\mathbf{A})}, \mathbf{e}_{1}^{(\mathbf{C})}\big ]$ (although this only occurs for 0.5 + 1.8 = 2.3\% tensors). Because this region accounts for 38.6\% of the data, strong signs of axisymetric extension are required elsewhere in $\mbox{Q}-\mbox{R}$ space for this state to be so prevalent. Consequently, the contribution by $\big [\mathbf{e}_{3}^{(\mathbf{A})}, \mathbf{e}_{3}^{(\mathbf{C})}\big ]$ when $\mbox{R} < 0$ (24\% of all the cases) is crucial, particularly when $\mbox{Q} > 0$ (17.4\%). Indeed, if it were not for this alignment state, the $e_{LR}^{(\mathbf{A})} = 1$ pattern would really be a Vieillefosse tail phenomenon, only. The dominance of the $\big [\mathbf{e}_{3}^{(\mathbf{A})}, \mathbf{e}_{3}^{(\mathbf{C})}\big ]$ alignment in this region also explains the large positive enstrophy production (the only part of the $\mbox{Q}-\mbox{R}$ space where this occurs so strongly\cite{tsinober01}) and weak strain production found here. This is shown in the first row of Fig. \ref{fig.sss_wws}. 

While no other alignment dominates a region of $\mbox{Q}-\mbox{R}$ space to the same extent as $\big [\mathbf{e}_{3}^{(\mathbf{A})}, \mathbf{e}_{3}^{(\mathbf{C})}\big ]$ when $(\mbox{R} < 0, \mbox{Q} > 0)$ , the second most dominant occurence of all is when $(\mbox{Q} > 0$, $\mbox{R} > 0)$, and involves $\big [\mathbf{e}_{1}^{(\mathbf{A})}, \mathbf{e}_{1}^{(\mathbf{C})}\big ]$. These data are shown in the third row of Fig. \ref{fig.sss_wws} and show very different behavior to the other $\mbox{R} > 0$ cases shown, with very strong negative enstrophy production driving the response, rather than positive strain production. The other $\mbox{R} > 0$ cases highlighted include those involving second eigenvector alignments, which emerge as important near the Vieillefosse tail, with $\big [\mathbf{e}_{2}^{(\mathbf{A})}, \mathbf{e}_{2}^{(\mathbf{B})}\big ]$ more important as one approaches from below $(\mbox{D} < 0)$, and $\big [\mathbf{e}_{2}^{(\mathbf{A})}, \mathbf{e}_{2}^{(\mathbf{C})}\big ]$ from above $(\mbox{D} > 0)$.

From the perspective of Taylor's result that $\langle\omega_{i}\omega_{j}S_{ij}\rangle > 0$ \cite{taylor37}, our results clearly permit the alignments that act counter to this average behavior to be discerned, and the case where this arises irrespective of position on the $\mbox{Q}$-axis when $\mbox{R} > 0$ is $\big [\mathbf{e}_{1}^{(\mathbf{A})}, \mathbf{e}_{1}^{(\mathbf{C})}\big ]$. The fifth and fourth rows also show that this situation occurs where $\big [\mathbf{e}_{3}^{(\mathbf{A})}, \mathbf{e}_{3}^{(\mathbf{B})}\big ]$ dominates for $\mbox{D} > 0$, and 
$\big [\mathbf{e}_{2}^{(\mathbf{A})}, \mathbf{e}_{2}^{(\mathbf{B,C})}\big ]$ for $\mbox{Q} > 0$.

The second most important alignment when $\mbox{R} < 0$, is $\big [\mathbf{e}_{1}^{(\mathbf{A})}, \mathbf{e}_{1}^{(\mathbf{B})}\big ]$, which grows from 4\% of cases for $\mbox{Q} > 0$, through 22\% for $(\mbox{D} > 0, \mbox{Q} < 0)$, to 46\% for $\mbox{D} < 0$. The first two rows of Fig. \ref{fig.sss_wws} show an important difference with $\mbox{R} < 0$ arising because of positive enstrophy production in both cases, but $-S_{ij}S_{jk}S_{ki}$ counteracting positive enstrophy production in the first row, and negative strain production acting with positive enstrophy production in the second row. Indeed, this is a consistent property: if the dominant alignment is between $\mathbf{S}_{\mathbf{A}}$ and $\mathbf{S}_{\mathbf{B}}$ there is a `consistent' behavior for predicting the sign of $\mbox{R}$ from the production terms. In contrast more refined analysis of the balance of these terms is needed when the dominant alignments are between $\mathbf{S}_{\mathbf{A}}$ and $\mathbf{S}_{\mathbf{C}}$. Hence, while the dynamical significance of separating $\mbox{R}$ into its constituents has already been shown \cite{luthi09}, a new rationale for this may be offered here: When dominant strain alignments are between $\mathbf{S}_{\mathbf{A}}$ and $\mathbf{S}_{\mathbf{C}}$  the sign of $\mbox{R}$ does not determine the nature of strain and enstrophy production in any simple way. Because the eigenvalues for $\mathbf{A}$ and $\mathbf{B}$ are identical, their behavior in $\mbox{Q}-\mbox{R}$ is also identical. Consequently, disaggregating into the constituent terms is essential for revealing the contribution of $\mathbf{C}$ to the fluid mechanics of trajectories within the space of the reduced Euler approximation.

\section{Properties of the $\kappa$ decomposition}
Mathematical constraints on the values for $\kappa_{2}$ and $\kappa_{3}$ in (\ref{eq.kappa12}) and (\ref{eq.kappa3}) simplify our analysis to two planes: 
\begin{itemize}
\item A $\kappa_{1}-\kappa_{3}$ plane if $\mbox{D} < 0$ (with $\kappa_{3} = \pm 1$ for $\mbox{D} > 0$; $\kappa_{3} = -1$ if $\mbox{R} < 0$, $\kappa_{3} = 1$ if $\mbox{R} > 0$); and, 
\item A $\kappa_{1}-\kappa_{2}$ plane if $D > 0$ (with $\mbox{D} < 0$ values at $\kappa_{2} = -1$). 
\end{itemize}
The $\kappa_{2}$ term has the property that $0 < \kappa_{2} \le 1$ corresponds to $\mbox{Q} > 0$ and $-1 < \kappa_{2} < 0$ corresponds to $\mbox{D} > 0$ and $\mbox{Q} < 0$. That is, the discriminant function in $\mbox{Q}-\mbox{R}$ space is recast here as the relative magnitudes of the Frobenius norms for the symmetric and skew-symmetric parts of $\mathbf{B}$. The corresponding constraint for $\kappa_{3}$ is that $0 < \kappa_{3} \le 1$ equates to $\mbox{R} > 0$ and $-1 < \kappa_{3} < 0$ to $\mbox{R} < 0$. 

\begin{figure}[t]
\vspace*{2mm}
\begin{center}
\includegraphics[width=14.2cm]{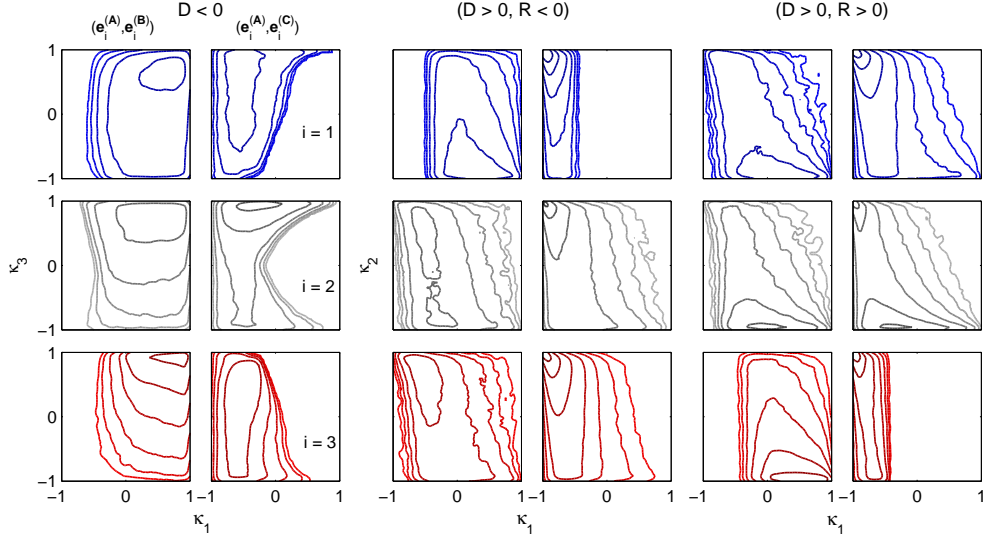}
\end{center}
\caption{Contour plots of the joint probability of $\kappa_{1}-\kappa_{3}$ for $\mbox{D} < 0$ and $\kappa_{1}-\kappa_{2}$ for $\mbox{D} > 0$ as a function of the strain alignments also used in Fig. \ref{fig.eLR_A}. These latter results are sub-divided by the sign of $\mbox{R}$. Each two-column set of results shows the strain alignments for $(\mathbf{e}_{i}^{(\mathbf{A})},\mathbf{e}_{i}^{(\mathbf{B})})$ (left-hand column) and $(\mathbf{e}_{i}^{(\mathbf{A})},\mathbf{e}_{i}^{(\mathbf{C})})$ (right-hand column). The blue, gray, and red contours are for $i = 1$, $i = 2$, and $i = 3$, respectively. The contour lines are distributed on a logarithmic scale from $10^{-5}$ to $10^{-2}$ in half-integer intervals of the power of ten.} 
\label{fig.jointkappa}
\end{figure}            

As the dominant alignments are between $[\mathbf{e}_{i}^{(\mathbf{A})}, \mathbf{e}_{i}^{(\mathbf{B})}\big ]$ when $\mbox{D} < 0$ (Fig.\ref{fig.eLR_A}), we find $\kappa_{1}$ is typically positive for these cases in the $\kappa_{1}-\kappa_{3}$ plane. Given that $\kappa_{3}$ is the Lund and Rogers normalization of $\mathbf{S}_{\mathbf{B}}$, this positive tendency is indicative of a dominant axisymmetric, expansive straining structure for $\mathbf{B}$ in the $\mbox{D} < 0$ region.

When $\mbox{D} > 0$ the behavior of $[\mathbf{e}_{i}^{(\mathbf{A})}, \mathbf{e}_{i}^{(\mathbf{B})}\big ]$ depends on both the sign of $\mbox{R}$ and that of $\mbox{Q}$: first eigenvector alignment when $\mbox{R} < 0$ (third column) and third when $\mbox{R} > 0$ (fifth column) gives $\kappa_{1} > 0$. Otherwise, $\mbox{Q} > 0$ (upper half of the $\kappa_{1}-\kappa_{2}$ planes) gives a more negative response, which dominates the distribution functions when $\mbox{R} < 0$. In contrast, when $[\mathbf{e}_{i}^{(\mathbf{A})}, \mathbf{e}_{i}^{(\mathbf{C})}\big ]$ alignments are dominant, $\kappa_{1}$ is strongly negative. The exception to this are the $[\mathbf{e}_{2}^{(\mathbf{A})}, \mathbf{e}_{2}^{(\mathbf{C})}\big ]$ cases in the vicinity of the Vieillefosse tail (upper half of the panel in the middle row, second column; lower half of the of the panel in the middle row, sixth column). This highlights a property of straining in this region that differs from other cases and is not discernible in the production terms shown in Fig. \ref{fig.sss_wws}.

\section{Conclusion}
While consideration of turbulence in a Fourier shell representation views dissipation as a small-scale phenomenon, when studying spatial fields at small-scales (near to Kolmogorov scales), where velocity derivatives dominate, all locations are `small'. Hence, focus moves toward the locations where particular phenomena occur, leading to the increasing focus on the topological approach \cite{chong90,meneveau11,zhou14} to understand processes such as dissipation \cite{goto08,goto09}. Here we have proposed a new representation of the topological space for these phenomena based on a decomposition of the velocity gradient tensor into normal and non-normal components. Because the eigenvalues of the former are equal to those of the velocity gradient tensor, their $\mbox{Q}-\mbox{R}$ representations are equivalent. Hence, explicit consideration of non-normality provides a richer representation, complementing the pre-existing suggestion to separate enstrophy production and strain production terms \cite{luthi09}. 

Enhanced closure models for engineering models of turbulent fluid flow are likely to be based on knowledge of Lagrangian walks around such topology spaces \cite{martin98,ooi99}. Our $\kappa_{1}-\kappa_{2,3}$ planes provide a new way to formulate such models. An extension of the current work is therefore an examination of the Lagrangian dynamics in $\kappa_{1}-\kappa_{2,3}$ space, perhaps conditioned on the strain alignments that have been shown to be of dynamical importance here. Another area to explore would be the extension of the reduced Euler system to incorporate pressure Hessian effects in the context of our approach \cite{wilczek14}. Such refinements would provide an alternative means to describe spatially intermittent dissipation in a modeling context \cite{horiuti16}.

\bibliography{PRLk}

\end{document}